\documentstyle[aas2pp4,psfig]{article}
\newcommand{\etal}{{{ et al.}}~}
\newcommand{\eg}{{{e.g.,}}~}

\newcommand{\kms}{{{km s$^{-1}$}}~}
\newcommand{\kmsc}{{{km s$^{-1},$}}~}
\newcommand{\kmsp}{{{km s$^{-1}.$}}~}
\newcommand{\Ho}{{{H$_\circ$}}~}
\newcommand{\amin}{{{$^\prime$}}~}
\newcommand{\asec}{{$^{\prime\prime}$}~}

\begin{document}

\title{\bf A PREDICTION OF OBSERVABLE ROTATION \\IN THE ICM OF ABELL 3266}

\vspace{1.in}
\author{KURT ROETTIGER}
\affil{Department of Physics and Astronomy\\University of
Missouri-Columbia\\Columbia, MO 65211\\E-mail:
kroett@hades.physics.missouri.edu}
\author{RICARDO FLORES}
\affil{ Dept. of Physics and Astronomy\\University of Missouri-St. Louis\\St.
Louis, MO 63121-4499\\E-mail: Ricardo.Flores@umsl.edu}

\vspace{.5in}

\begin{center}{Accepted for publication in the Astrophysical Journal}
\end{center}




\begin{abstract}
We present a numerical Hydro+N-body model of A3266
whose X-ray surface brightness, temperature distribution,
and galaxy spatial and velocity distribution data
are consistent with the A3266 data. The model is
an old ($\sim$3 Gyr), 
off-axis merger having a mass ratio of $\sim$2.5:1.
The less massive subcluster in the model
is moving on a trajectory from
southwest to northeast passing on the western side of the dominant
cluster while moving into the plane of the sky at $\sim$45 degrees.
Off-axis mergers such as this one
are an effective mechanism for transferring angular momentum
to the intracluster medium (ICM),
making possible a large scale rotation
of the ICM. We demonstrate here that the  ICM rotation predicted by our
fully 3-dimensional  model of A3266 is observable with current technology.
As an example, we present simulated observations assuming the 
capabilities of the high resolution X-ray spectrometer (XRS) which
was to have flown on {\it Astro-E}.

\end{abstract}


\section{Introduction}

A3266 is a nearby (z=0.059; Quintana, Ramirez \& Way 1996),
X-ray luminous cluster which exhibits
optical and X-ray substructure. Two models have recently been proposed to
explain the substructure in this cluster. Flores, Quintana, \& Way (1999)
have proposed that A3266 experienced a major merger {\it into} the plane of the sky while Henriksen, Donnelly, \& Davis (1999) have proposed a minor
merger parallel to the plane of the sky.
Using a large sample of galaxy redshifts (387 galaxies), Quintana \etal (1996)
suggested that A3266
might have
experienced a merger 1-2 Gyr ago. Flores et al. (1999) found support for
this interpretation using a simple N-body model.
They noted an enhancement of galaxies north of the X-ray core
similar, both visually and statistically, to
the N-body particle spray found in their numerical simulations.
They also noted
a similar enhancement of emission-line galaxies in the same region.
It has long been suggested that galaxies passing through cluster cores
could be spectroscopically altered (Dressler \& Gunn 1983), although it now
appears that this would be mostly due to the tidal force rather than ram
pressure by the ICM (\eg Moore \etal 1996; Bekki 1999; Fujita \etal 1999). Burns
\etal (1994) have suggested that the Coma cluster E+A galaxies distributed in
the core and SW toward the NGC 4839 group
are the result of a burst of star formation induced by a merger about 2 Gyr ago,
which appears consistent with their starburst age (Caldwell \etal 1996). Similarly, the emission-line galaxies in A3266 could be the tail end of the disrupted less
massive cluster in this model.

Evidence of recent dynamical evolution is also
apparent in the X-ray surface brightness (XSB; Fig. \ref{a3266})
and temperature distributions. For example, the XSB was shown to exhibit a
systematic centroid shift by Mohr, Fabricant \& Geller (1993). Also, the
XSB exhibits changing ellipticity and isophotal twisting between 4\amin and
8\amin (Mohr \etal 1993) as well as a large ($\sim$500 kpc) core radius
(Mohr, Mathiesen \& Evrard 1999), as would be expected in the case of a recent
merger (Roettiger \etal 1996).
Peres \etal (1998) find no evidence of
a cooling flow. Several researchers (\eg McGlynn \& Fabian 1984) have
suggested
that mergers will disrupt cooling flows. G\'omez \etal (1999) have
demonstrated using numerical simulations
that the time scale for re-establishment of the disrupted cooling flow
in the post-merger environment can be greater than several billion years,
depending on the initial cooling flow and merger parameters. Markevitch \etal
(1998; hereafter MFSV98)
and Henriksen \etal (1999; hereafter HDD99) have  produced
temperature
maps based on {\it ASCA} data that show significant temperature variations across
the cluster. MFSV98 shows a radially decreasing temperature profile 
ranging from 12 keV in the central 3\amin to $\sim$6 keV at radii greater than
8\amin. De Grandi \& Molendi (1999) find a similar radial temperature gradient using {\it BeppoSAX} data. The HDD99 temperature map exhibits
a comparable range in ICM temperatures with a gradient increasing from NE to SW
across the cluster. 

In this paper, we extend the N-body model of Flores \etal (1999) by including the
hydrodynamics of the ICM. We then demonstrate using a fully 3-dimensional numerical
model that the current A3266 data are consistent with an old off-axis merger
occurring largely {\it into} the plane
of the sky.  Off-axis mergers are
a natural consequence of large-scale tidal 
torques, the latter being
a generic feature of hierarchical clustering
(Peebles 1969).
The model of Flores et al. (1999) resulted in an off-axis merger as a result
of the global angular momentum imposed at the protocluster stage, which in
terms of the standard dimensionless angular momentum $\lambda$ (Peebles 1969)
corresponded to $\lambda = 0.07$. This amount is consistent with tidal torquing
and is expected to be largely independent of mass (Barnes \& Efstathiou 1987).
Therefore the characteristics of the merger are not sensitively dependent on
extending the region that was simulated around the cluster.
In sufficient
quantity, angular momentum can significantly alter the internal structure
of clusters, which can then influence our interpretation of other cluster
observations. As an example, numerical simulations (Inagaki \etal 1995;
Roettiger \etal 1997) have
shown that the
shape of clusters (\eg oblateness being a consequence of rotation) can have
significant systematic effects on determinations of H$_\circ$ based
on the Sunyaev-Zeldovich effect (see Birkinshaw (1999) for a review).

 We describe our model and make direct comparisons to the data in
\S\ref{model}.
In \S\ref{strategy}, we present detailed models of proposed
{\it Astro-E} observations based on the line-of-sight (LOS) ICM density,
temperature
and velocity structure provided by the numerical model.  Section \ref{summary}
is
a summary of our results. We assume \Ho=70 \kms Mpc$^{-1}$ when scaling the 
simulation to physical dimensions.

\section{A Numerical Model of A3266.}
\label{model}
We have created a numerical model of A3266 using 
the same technique that we have employed in several previous models 
of specific Abell clusters
(\eg A2256 Roettiger \etal 1995; A754, Roettiger \etal 1998;
A3667, Roettiger, Burns \& Stone 1999).  Within the framework of idealized initial 
conditions, we survey merger parameter space (mass ratios, impact parameters,
gas content, etc.) using a fully 3-dimensional Hydro/N-body 
code based on the Piecewise-Parabolic Method (PPM; Colella \& Woodward 1984)
and
a Particle-Mesh (PM) N-body code. 
We then attempt to maximize 
agreement between synthetic observations of the simulation and various cluster
observables (X-ray surface brightness, X-ray temperature distribution, 
galaxy spatial and velocity distributions etc.) in order to constrain
not only the merger parameters, but also the epoch and
viewing geometry of the merger.  Here, we have modeled A3266
as an off-axis merger between a primary cluster 
of $\sim$1.1$\times$10$^{15}$M$_\odot$
and a secondary of $\sim$5$\times$10$^{14}$M$_\odot$ in which
closest approach occurred approximately 3 Gyr ago. The secondary cluster, moving
southwest to northeast, passed to the west of the primary cluster's core at a
distance
of $\sim$230 kpc with a velocity of $\sim$2500 \kmsp The trajectory is believed 
to be into the plane of the sky at an angle of $\sim$45$^\circ$.

Figure \ref{a3266sim} shows the synthetic XSB image 
generated from the model. Like the {\it ROSAT} PSPC image (Fig. \ref{a3266}), the simulated image
shows a generally spherical distribution at large radii with significant
isophotal twisting near the X-ray core which is elongated NE to
SW. The orientation of the X-ray cores (both simulated and observed)
are not well-aligned
with the projected mass distribution derived from gravitational
lensing (Joffre \etal 1999) further indicating that the cluster is not
fully relaxed. It should be noted that the resolution of the numerical
simulations ($\sim$75 kpc or $\sim$4 zones core radius) is significantly less
than the resolution of the {\it ROSAT} image (15\asec/pixel or
$\sim$20$h^{-1}_{70}$ kpc).

Also included in Fig. \ref{a3266sim} is a sampling  of $\sim$300 N-body
particles ($<$1\% of total) from both the primary ($\diamond$) and secondary (+) clusters. 
The excess of secondary cluster particles to the north of the X-ray core
accounts for the galaxy excess as well as the distribution of emission-line
galaxies
noted by Flores \etal (1999). The observed galaxy
velocity distribution is indistinguishable from Gaussian within
the central region.
The observed skewness and kurtosis within the central 1$^\circ \times 1^\circ$
field are 0.024 and 0.14, respectively, while the N-body particle
values are  0.051$\pm$0.15 and 0.13$\pm$0.26. On a larger  scale ($\sim$2$^\circ$), Flores \etal
(1999)
find the observed skewness and kurtosis to be 0.106 and 0.341, respectively. The global N-body
velocity dispersion is 905$\pm$40 \kmsc while Quintana \etal (1996)
observe a global galaxy velocity dispersion of 1085$\pm$51 \kmsp The
discrepancy
here is possibly due to choosing too low a value for the initial cluster
$\beta$-parameter.

A3266 contains a central dumb-bell galaxy with a velocity separation of $\sim$400 \kmsp 
Quintana \etal (1996) suggest that this is consistent with the merger geometry proposed here. 
Although consistent with a younger merger that is nearly in the plane of
the sky, it is also consistent with an older merger at any projection since the time scales for dynamical friction
to act on the dominant galaxy are long (Kravtsov \& Klypin 1998). The velocity
separation noted here should be contrasted with the 2635\kms separation observed
between dominant galaxies in the proposed young merger A2255 (Burns \etal 1995).

It has been suggested that radio source morphology may give clues to
the ICM dynamics. The radio emitting plasma is believed to be 10 to 100 times less
dense than the surrounding thermal gas making it susceptible to pressure gradients
and flows within the ICM (\eg Burns 1998). HDD99 used the morphology of two radio sources
to support their model, and depending on the exact location of these sources
within the cluster, they are also consistent with the model presented here. We should
comment however that considerable caution must be used when
interpreting radio source morphology in this context.
Both sources in question
are located SW of the cluster core (see Jones \& McAdam 1992). One
appears to be a head-tail source while the other is identified as a possible 
Wide-Angle Tailed radio source (WAT). WATs have been used as indicators of
bulk flows
in clusters, because they have traditionally been associated with central
dominant
galaxies which are presumed to be at rest in the cluster's gravitational
potential minimum making knowable its exact location (and relative velocity) within the cluster. 
Consequently, any bending
of the WAT tails is the result of ICM dynamics rather than motion of the host
galaxy.
However, both of these radio sources are associated with galaxies having
velocities
significantly different ($\Delta$V$\sim$800 \kms)  from the mean cluster velocity
indicating
that they may have considerable velocities of their own and may even be
foreground or background objects.

Figure \ref{a3266simt} shows the projected, emission-weighted temperature
map overlaid with the XSB contours.
Often the X-ray temperature distribution provides the strongest constraints
on the merger parameters. Two X-ray temperature maps have been published
recently
based on {\it ASCA} data (MFSV98; HDD99). Although
largely consistent, they do differ systematically. Both maps show a similar
degree of temperature variation ($\sim$6-12 kev) within the cluster. However,
the MFSV98 map shows a hot core with a radially decreasing temperature
profile while the HDD99 map shows more of a temperature gradient
across the cluster. The core is not the hottest region in the HDD99 map. 
For this reason,
we have performed a region-by-region comparison of our model with both 
published temperature maps. Figure \ref{tcomp}a is a comparison between 
our model and the corresponding
regions in the MFSV98 map. Figure \ref{tcomp}b compares the model
temperatures within regions defined by HDD99 (see Fig. \ref{a3266simt} for region definition). 
Our model agrees within the 90\% confidence intervals of all but one  region defined 
by MFSV98. In an absolute sense, the agreement with the HDD99 map is just as
good
although their quoted uncertainties are significantly smaller than those
quoted
by MFSV98. The most significant discrepancy between our model and HDD99 is
in region 5 which includes the cluster core. Although our model agrees
quite well with MFSV98 in this region (Region 1; Fig. \ref{tcomp}a), the HDD99
map 
indicates a cooler core. If correct, this could be an indication that
cooling may have started to influence the core (where cooling
times are the shortest) because the merger is relatively old.
Our current model does not include the effects of radiative cooling.

A3266 has recently been observed in the hard X-ray band (15-50 keV) 
by {\it BeppoSAX} (De Grandi \& Molendi 1999). They found no evidence of a hard
X-ray component nor is there currently any evidence of  diffuse radio emission. This
result is consistent with an old merger interpretation since shocks present
in young mergers might be expected to accelerate relativistic particles
which may produce observable
diffuse radio emission via synchrotron and hard X-rays via inverse Compton scattering (see Sarazin 1999
for a recent review).

Given the model presented here, what can we say about the gas dynamics
in A3266? First, A3266 appears to be an old off-axis merger in which
the angular momentum of the initial clusters is due to tidal torquing 
(Flores \etal 1999). 
Previous off-axis merger simulations (Roettiger \etal 1998; Ricker 1998) have
shown that
mergers can transfer significant angular momentum to the ICM and that this
angular momentum can be long-lived. In fact, because of the time and distance scales
involved,
it takes a considerable period (several billion years) for full rotation to develop.
To understand this, consider a mean rotational velocity, $v_{rot}$=1000 \kmsc
at a radius, $r=$500 kpc. The time required to complete circulation about the core
is then simply $t=2\pi r/v_{rot}$ or $\sim$3 Gyr. In this model, full rotation
occurs between 2.5 and 3 Gyr after core passage and persists beyond 4 Gyr. Of
course this is in the absence of a second significant merger event which could
potentially disrupt the circulation.

Figure \ref{simrot} shows the gas velocities within an east-west plane taken
along
the observer's LOS and through the cluster core. Note that even at
this late stage of the merger, there are still significant gas velocities
($>$500 \kms) 
and that full rotation has been established.  Further note, that shocks
generated early in the merger have now dissipated while extremes in the
temperature distribution have not. As we
have pointed out previously (Roettiger \etal 1996), substructure of this
type can persist well beyond the
canonical sound crossing time. Thus, it
may significantly influence estimates
of cosmological parameters based on
the frequency of substructure in clusters (\eg Richstone,
Loeb, \& Turner 1992).
As we will see in \S\ref{strategy}, our model not only predicts
fully developed rotation in A3266, but also predicts a viewing geometry that 
places a significant component of the opposing bulk flows along the observer's
LOS, thus enhancing the prospect of detecting it spectroscopically.

\section{Observing the ICM Rotation with {\it Astro-E}}
\label{strategy}

The {\it Astro-E} XRS (Audley \etal 1999) is a high resolution X-ray
spectrometer scheduled for
launch in early 2000. With $\sim$10 eV energy resolution and a quantum efficiency near
unity
across the 0.4 to 10 keV energy band, it represents the first opportunity to
directly 
observe gas dynamics in the ICM. Of course, these observations will be limited
to
LOS gas velocities, and will be confused by multiple temperature, velocity,
and
possibly metal abundance components along the LOS. For this reason, it is extremely
useful
to have a 3-dimensional model of the gas density, temperature, and dynamical
structure 
both when planning observations and interpreting the data.

Figure \ref{simvmap} represents a simple LOS emission-weighted mean velocity
map
based on the numerical model. Analogous to the temperature map (Fig.
\ref{a3266simt}),
velocities within a given computational zone along the LOS are simply weighted
by the 
X-ray volume emissivity ($\propto n^2 T^\frac {1} {2}$). The range in
velocities across the 
cluster is greater than 800 \kmsp The opposing bulk motions evident on either
side of
the X-ray core constitute a strong signature of ICM rotation. West of the core,
gas
is moving away from the observer at greater than 500 \kms while east of the
core, gas
is moving toward the observer at greater than 300 \kmsp Of course, this
velocity
map is only indicative of the observable LOS gas motions in the cluster. 
Below,
we discuss detailed simulations of {\it Astro-E} observations.

Based on the model, we propose two pointings selected to optimize their expected
velocity separation and  XSB.
We suggest one pointing on each side of of the X-ray core,
separated by $\sim$8\amin ($\sim600$$h_{70}^{-1}$) on an east-west line
running 2\amin north of the X-ray maximum. The pointing location is
chosen in part to avoid the point source located $\sim$4\amin due east of the
core.
The expected mean emissivity-weighted velocities at these two
locations are -195 \kms (east) and +533 \kms (west) for a $\Delta$V=728 \kmsp

In order to test the feasibility of this observation, we have used our
numerical model and
XSPEC (Arnaud 1996) to generate synthetic X-ray spectra for each of the two
pointings described above. 
Each spectrum is a composite of 50 spectra 
(Raymond-Smith+absorption) characterized
by the local ICM temperature, density, velocity, and chemical abundance within
a computational zone
along the observer's LOS. We have assumed a uniform abundance of 0.2 solar (De
Grandi \& Molendi 1999).
The total flux  of the simulated cluster is scaled to
2.84$\times$10$^{-11}$ erg s$^{-1}$
 cm$^{-2}$ (0.5-2.0 keV; David \etal 1999) while the absorption is characterized by
N$_h$=3.0$\times$10$^{20}$
cm$^{-2}$ (White \etal 1999). De Grandi \& Molendi (1999) quote a somewhat lower absorption
of 
N$_h$=1.6$\times$10$^{20}$cm$^{-2}$. Using the {\it ROSAT} PSPC image as a flux
distribution
template, a list of photon events is generated using MKPHLIST\footnote
{http://heasarc.gsfc.nasa.gov/docs/astroe/}. Optimum
exposures are determined
to be 50 ksec (east) and 60 ksec (west). The resulting events were then input
to XRSSIM$^1$. 
After excluding photons not in the XRS FOV, the
count rates for the east and west pointings are 0.20  and 0.17 cts s$^{-1}$,
respectively, or approximately 10$^4$ counts per spectrum. After rebinning of
the 
photons  to improve the statistics,
we produce the spectra
shown in Fig. \ref{simspec}. An isothermal fit 
to the 
Fe K-line complex (6.2-6.7 keV) yields a velocity centroid shift of -219$\pm$85 \kms east
of the core and +544$\pm$80 \kms west of the core for a velocity separation of 763$\pm$117 \kms
between the two pointings. These values are consistent with those expected
from a direct examination of the numerical model (see above), thus demonstrating that the ICM rotation
present in
the model when scaled to the XSB of A3266 is in fact
observable at a high level of significance with the {\it Astro-E} XRS.

\section{Summary}
\label{summary}

We have presented a 3-dimensional numerical Hydro +N-body model of A3266 which
is consistent with a wide range of observed properties. We believe A3266
represents
an old ($\sim$3 Gyr), off-axis merger that is occurring into the plane
of the sky at a 45$^\circ$ angle. The model is consistent, within resolution limits, with
the {\it ROSAT} PSPC
image, current {\it ASCA} temperature maps, {\it BeppoSAX} hard X-ray flux limits, 
the galaxy spatial and velocity
distributions, and the existing radio data. We have also checked that
the projected mass distribution
agrees with gravitational lensing data.

 In this model, the off-axis merger
has imparted significant angular momentum on the ICM of the merger
remnant which we predict should be observable with the XRS on {\it Astro-E}.
The signature of rotation will appear as two opposing bulk flows ($\Delta$V$\sim$800\kms) located 
on either side of the cluster's X-ray core along a line of constant declination.
The degree of circulation present in the core of our model is
consistent with high resolution Hydro+N-body simulations of galaxy cluster formation
from cosmological initial conditions (Norman \& Bryan 1998).
Neither the galaxy redshift data nor the N-body particle distribution
show similar evidence of rotation. Unlike the ICM interaction which
can be characterized as `collisional', the interaction between the subcluster
galaxy components is `collisionless'. Therefore, while the transfer
of angular momentum between ICM components is very efficient, angular
momentum is not transferred between the galaxy components.

It is difficult to assess the overall uniqueness of our model at the present time.
Individually, no
single
observation
places a strong constraint on the model. Even taken together, there is
considerable flexibility in the merger parameters and viewing geometry, and we
cannot rule out the possibility
that
recent mergers with significantly less massive subclusters 
have played some role (HDD99). Although
limitations in the current data set certainly contribute to this problem, it
is also indicative of an older merger that no distinctive features currently
exist. In the event ICM rotation is not detected, the
observations described here will nonetheless provide important
new constraints on the model and on ICM dynamics in general.
In addition, these data will result in highly accurate
temperature and metalicity measurements for two widely separated
regions within a single cluster thus giving clues to the degree
of spatial variation in both quantities. Inhomogeneities in the
distribution of metals may help quantify the rate of mixing between
gas components within the merger.

\acknowledgements
We would like to thank all the people who supported the {\it Astro-E}
mission which inspired this work. We would also like to express our deep regret at the loss of the instrument during launch.
We thank the Earth and Space Data Computing Division of NASA's Goddard Space Flight Center (GSFC) for use of
the MasPar2 on which these simulations were performed. We also thank the Laboratory
of High Energy Astrophysics at GSFC for making
available the {\it ROSAT} archival data through the W3Browse facility and
for making available the {\it Astro-E} simulation software. This
work has made use of NASA's Astrophysics Data System (ADS) Abstract Service
and the NASA/IPAC Extragalactic Database (NED). We thank
J. P. Henry and J. O. Burns for their useful comments and discussions.
The work of RF has been supported by a University of Missouri Research
Board Award. KR dedicates this work to the memory of George O. Minot
(1906-2000).


\clearpage

\figcaption[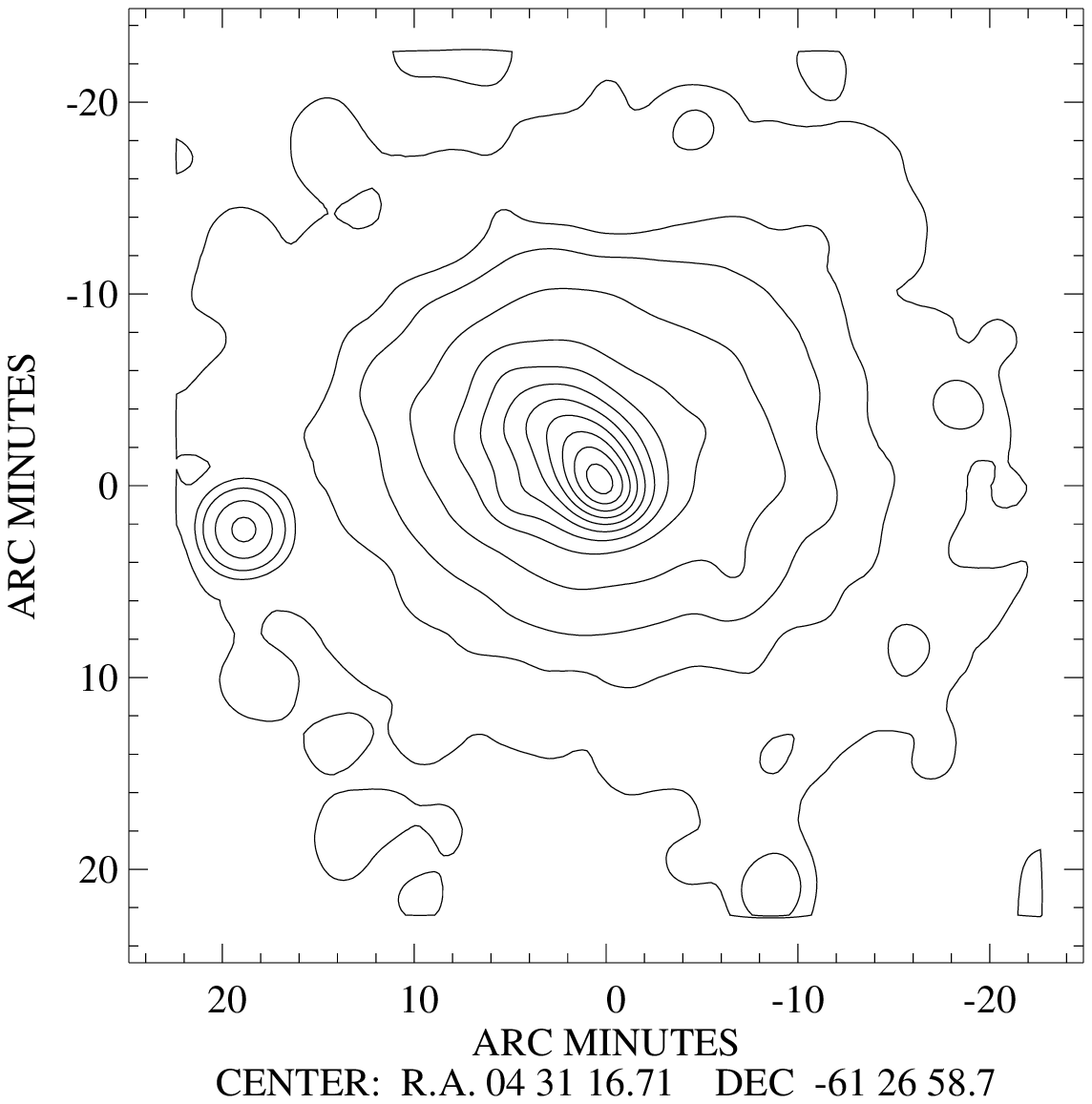]{The {\it ROSAT} PSPC archival image of Abell 3266. The image
is background subtracted. Contour levels are 0.02, 0.04, 0.07,
0.14, 0.25, 0.45, 0.55, 0.65, 0.75, 0.85, 0.95 of peak. \label{a3266}}

\figcaption[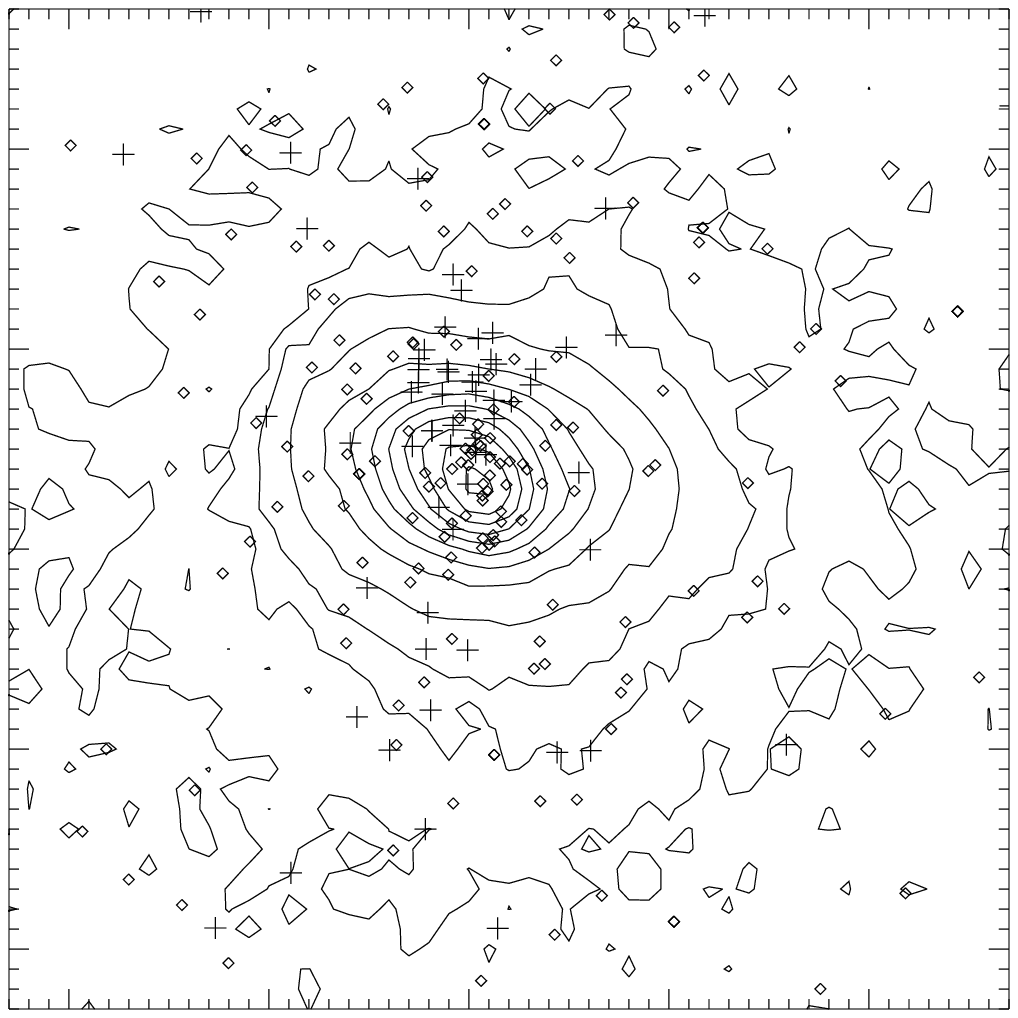]{ A synthetic X-ray surface brightness image
generated from the numerical model of A3266. Contours and linear
dimensions (assuming \Ho=70 \kms Mpc$^{-1}$) are the same as in 
Fig. \ref{a3266}. The $\diamond$'s and $+$'s represent primary and secondary cluster particles,
respectively. \label{a3266sim}}

\figcaption[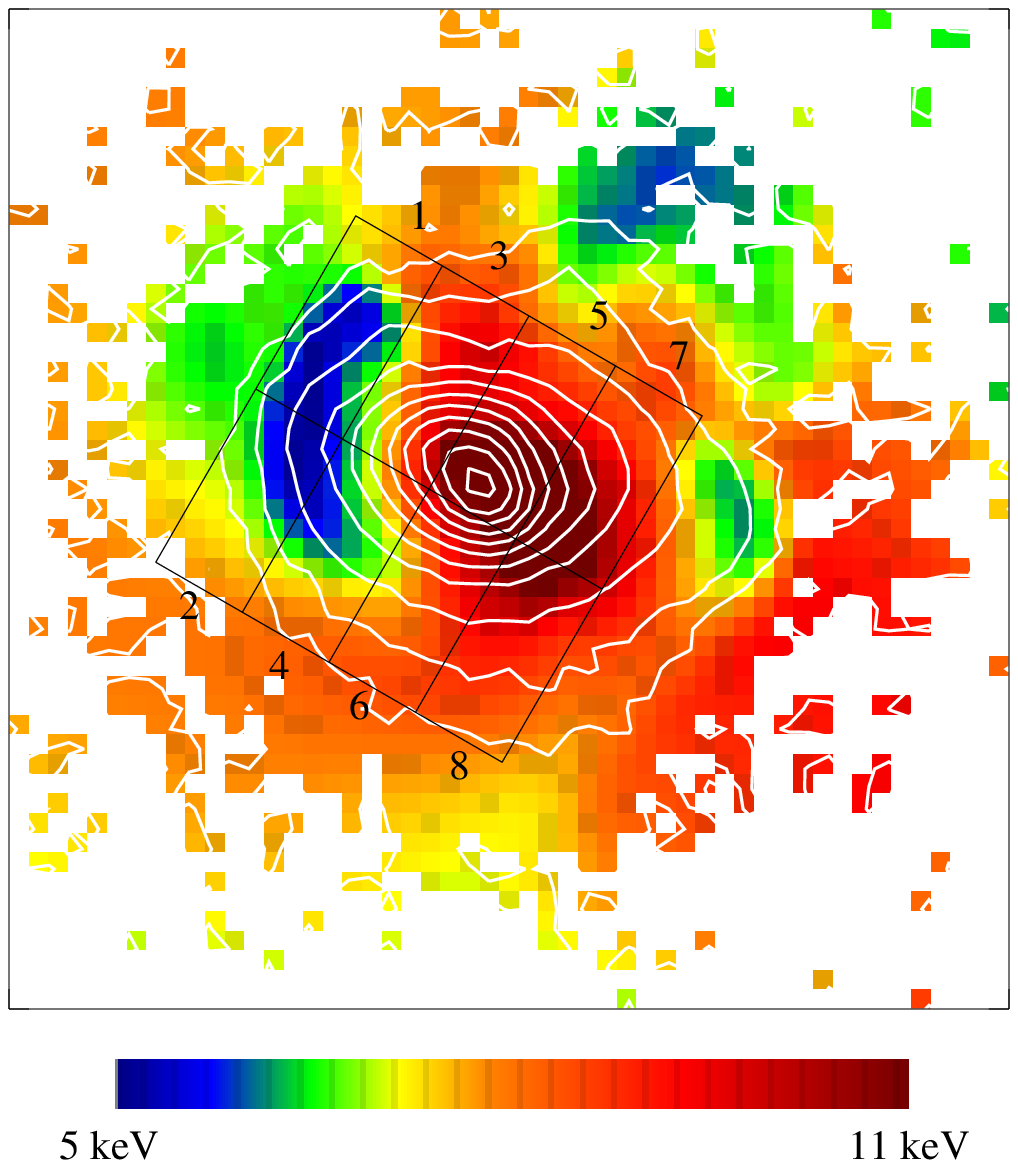]{ The projected emission-weighted temperature
map of the A3266 model (color) overlaid with the X-ray surface brightness
(contours). The numbered boxes are the temperature regions defined by HDD99.
\label{a3266simt}}

\figcaption[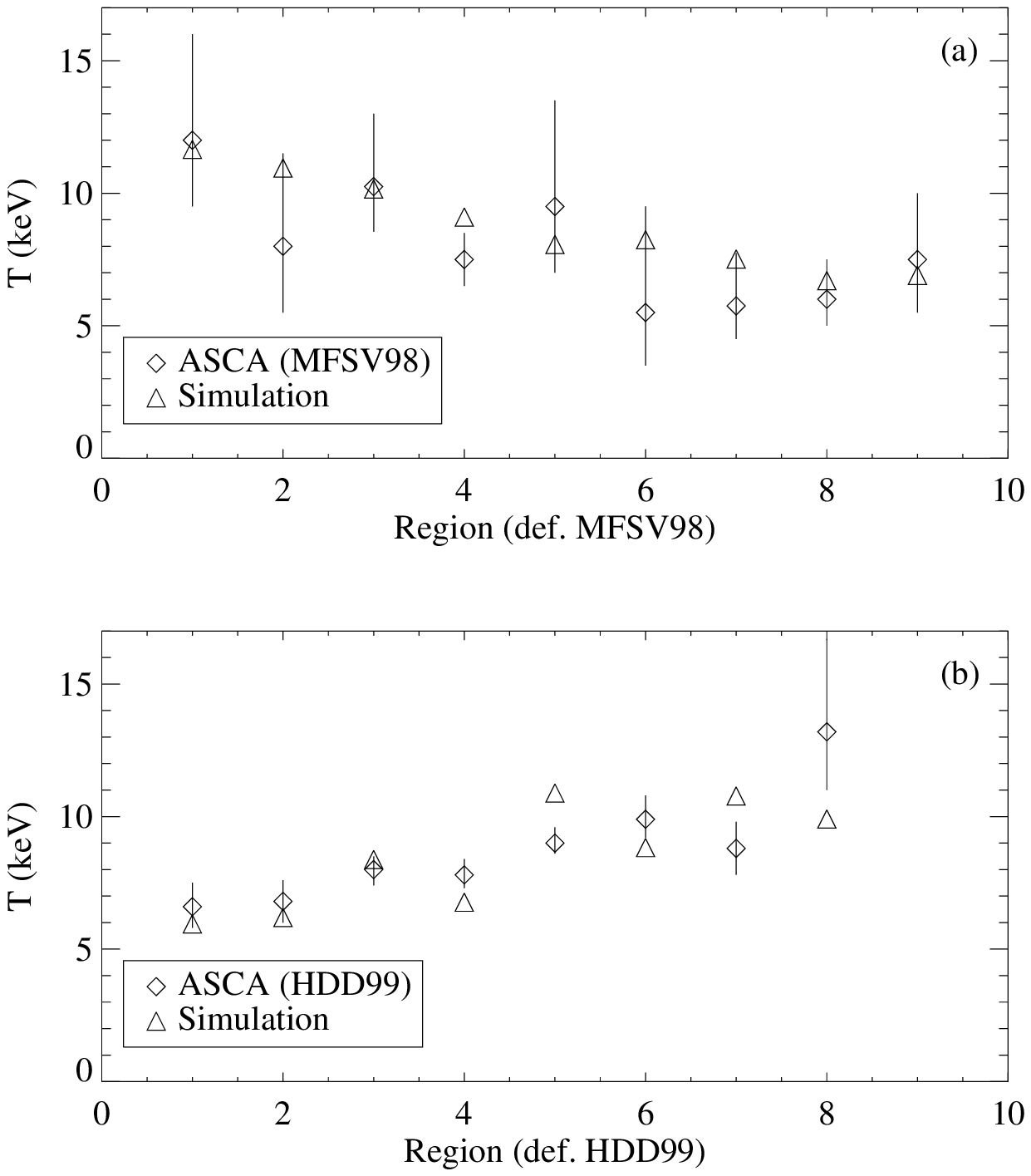]{Region-by-Region comparisons between the model
temperature distribution and that observed by a) MFSV99 and  b) HDD99
using {\it ASCA}. The regions defined by HDD99 are shown in Fig. \ref{a3266simt}.\label{tcomp}}
 
\figcaption[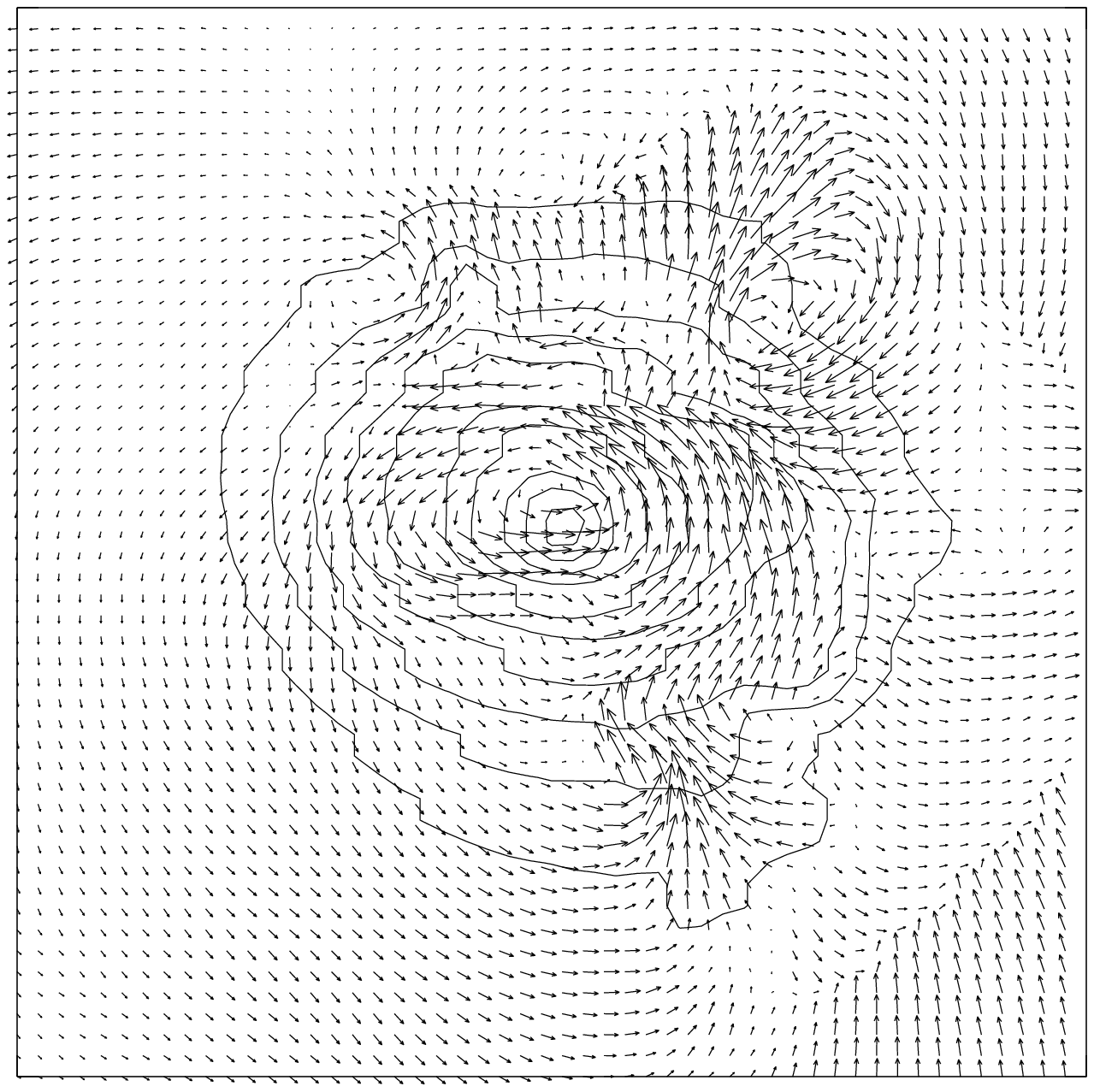]{ Velocity vectors 
overlaid with X-ray emissivity (contours)
within a 2-dimensional east-west (left to right) slice  taken through the
cluster core and
along the observer's LOS (bottom to top). 
Note the large-scale counterclockwise rotation near the cluster core.
Panel dimensions are 3.3 $\times$ 3.3 Mpc. The longest vector is $\sim$900 \kmsp
\label{simrot}}

\figcaption[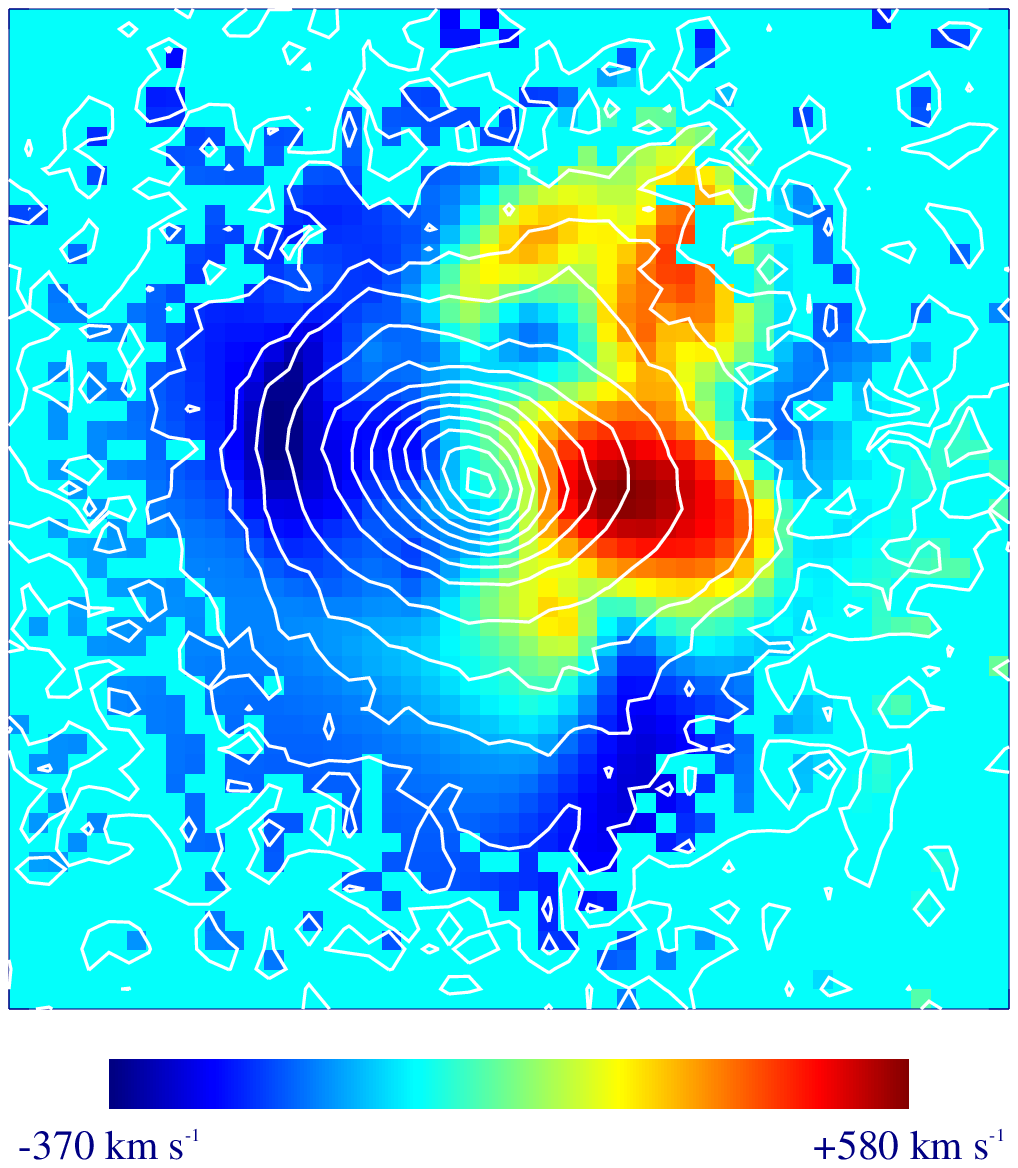]{Emission-weighted LOS ICM velocities
(color) overlaid with X-ray surface brightness (contours). Velocities
on the eastern half of the cluster (blue) are moving toward the observer
while velocities on the western half of the cluster (red) are moving
away from the observer. The background color (light blue) corresponds
to 0 \kmsp \label{simvmap}}

\figcaption[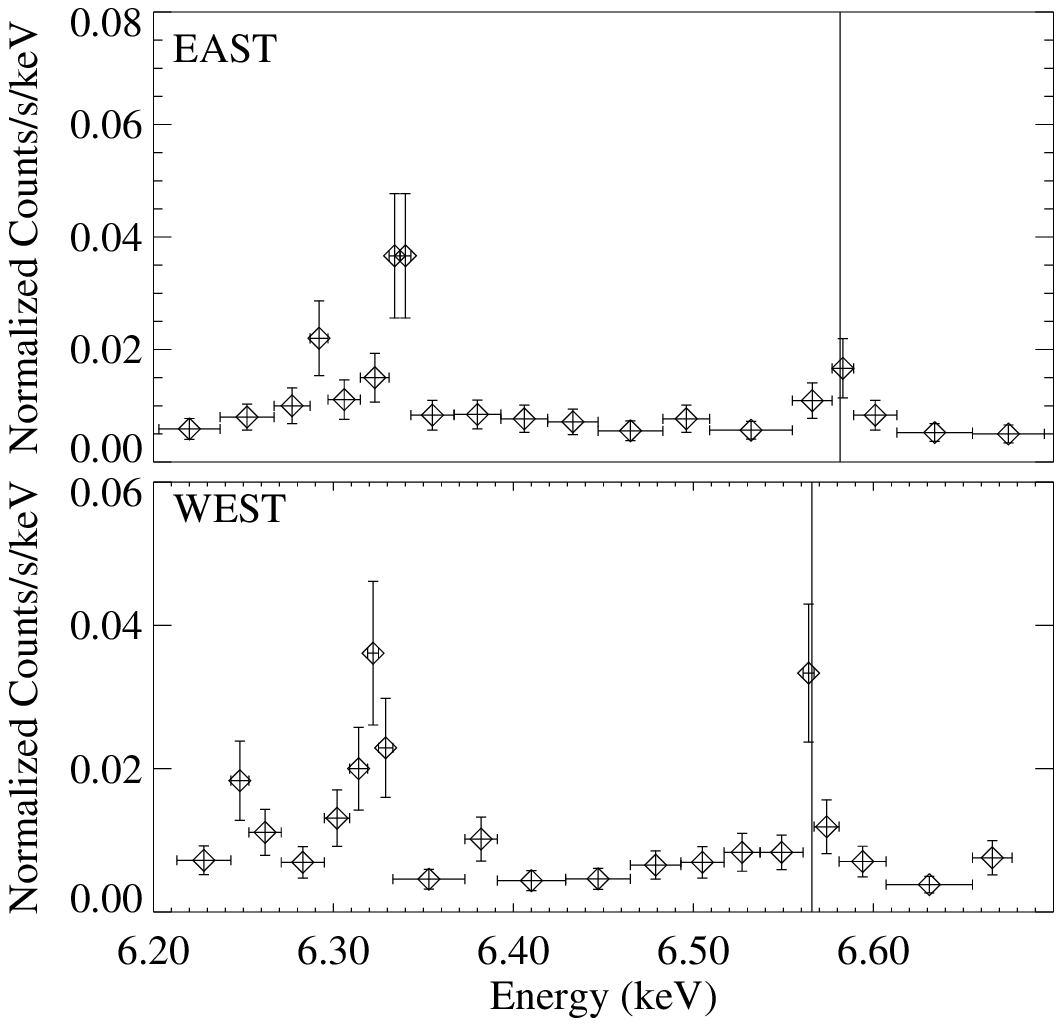]{ Simulated {\it Astro-E} spectra taken
4\amin on either side the X-ray core along an East-West line 2\amin
north of the core.
Each represents a 50-component model ($\rho$,T,$v$ from 50 zones within the
numerical simulation) normalized to the flux at the corresponding 
location in A3266. The vertical solid lines  indicate the location of the fit
to the line centroid. Integration times are 50 ksec East and 60 ksec West.
\label{simspec}}

\end{document}